\documentclass[preprint]{aastex}
\pdfoutput=1
\usepackage{times}
\usepackage{wasysym} % symbols
\usepackage{caption} %subfigures
\usepackage{subcaption} %subfigures
\usepackage{hyperref} % internal links
\usepackage{natbib} % bibliography
\usepackage{multirow}

\usepackage{graphicx}
\let\oldtabular\tabular
\renewcommand{\tabular}{\scriptsize\oldtabular}

\usepackage[normalem]{ulem}

\usepackage{color}

\begin{document}
\title{Metrics for Optimization of Large Synoptic Survey Telescope Observations of Stellar Variables and Transients}
\author{Michael B.\ Lund\altaffilmark{1},
Robert J.\ Siverd\altaffilmark{4},
Joshua A.\ Pepper\altaffilmark{2,1},
Keivan G.\ Stassun\altaffilmark{1,3}}
\altaffiltext{1}{Department of Physics and Astronomy, Vanderbilt University, Nashville, TN 37235, USA; \\ \url{michael.b.lund@vanderbilt.edu}}
\altaffiltext{2}{Department of Physics, Lehigh University, Bethlehem, PA 18015, USA}
\altaffiltext{3}{Department of Physics, Fisk University, Nashville, TN 37208, USA}
\altaffiltext{4}{Las Cumbres Observatory Global Telescope Network, Goleta, CA 93117, USA}

\captionsetup[table]{labelsep=space}
\captionsetup[figure]{labelsep=space}

\begin{abstract}
The Large Synoptic Survey Telescope (LSST) will be the largest time-domain photometric survey ever. In order to maximize the LSST science yield for a broad array of transient stellar phenomena, it is necessary to optimize the survey cadence, coverage, and depth via quantitative metrics that are specifically designed to characterize the time-domain behavior of various types of stellar transients. In this paper we present three such metrics built on the LSST Metric Analysis Framework (MAF) model \citep{Jones2014a}.  Two of the metrics quantify the ability of LSST to detect non-periodic and/or non-recurring transient events, and the ability of LSST to reliably measure periodic signals of various timescales. The third metric provides a way to quantify the range of stellar parameters in the stellar populations that LSST will probe. We provide example uses of these metrics and discuss some implications based on these metrics for optimization of the LSST survey for observations of stellar variables and transients.
\end{abstract}
%\maketitle
%\emph{Subject headings:} xxxxx --- xxxxx --- surveys

\section{Introduction}

The Large Synoptic Survey Telescope (LSST) is expected to provide time-domain photometric monitoring of billions of stars, over a period of $\sim10$ yr and with unprecedented breadth and depth. This will in turn permit an unprecedented ability to identify and characterize a variety of stellar variability and transient phenomena, from non-recurring novae and supernovae to periodic variables such as eclipsing binaries. It will also more generally probe a very broad array of stellar types and populations.

To more fully capitalize on the LSST potential to contribute to a broad range of astronomical interests, the LSST team has publicly released data and tools so that the wider community can contribute to discussions of how to optimize the potential of LSST on any topics outside of LSST's stated goals. The first of these tools is the LSST Operations Simulator (OpSim) which models the hardware and software performance, site conditions, and cadence as a function of sky position \citep{Delgado2014}. 
The other tool that has been provided by the LSST team is the Metric Analysis Framework (MAF)\footnote{Available at \url{https://confluence.lsstcorp.org/display/SIM/MAF+documentation}}. The objective of the MAF is to allow the comparison of different simulated LSST data sets by reducing the properties for each field or position in the sky to a single number which can then be mapped.  

In this paper, we introduce three MAF metrics\footnote{Available at \url{https://github.com/LSST-nonproject/sims_maf_contrib}} for use with LSST. These metrics are intended to address three questions:
\begin{itemize}
\item How well can a given observing plan detect %stationary {\bf [need a better word than stationary]}
transient photometric phenomena with constant equatorial coordinates?
\item With what level of completeness and precision can a given observing plan detect %stationary {\bf [again stationary]}
periodic photometric phenomena with constant equatorial coordinates? %May be better to just not try to include stationary in this description?
\item How many main sequence stars in the Milky Way will be detected in each specific LSST observing field, and what will be the range of their masses?
\end{itemize}

In \S \ref{sec:bg} we summarize the basic LSST specifications and simulation tools that we use.
In \S \ref{sec:metrics} we present the three metrics and describe how they can be used to characterize a given survey configuration (cadence, coverage, depth) toward optimization for various science applications. Then in \S \ref{sec:sum} we review the combined usefulness of the metrics presented here and describe how the development of further metrics can enhance the science yield of the LSST.

\section{ 
%Background and Context 
Basic LSST Specifications and Tools Used}
\label{sec:bg}

The Large Synoptic Survey Telescope will collect ten years of photometric observations in six bands over more than half the sky, scheduled to begin observations in 2020. LSST has the capability to contribute greatly to many areas of discovery, but there are four goals explicitly stated: Taking an Inventory of the Solar System, Mapping the Milky Way, Exploring the Transient Optical Sky, and Probing Dark Energy and Dark Matter \citep{LSSTScience2009}. The LSST system design has been dictated by these goals, and hardware specifications have already been established. The need to be able to survey the entire visible sky with relatively short (less than thirty seconds) exposures has resulted in a telescope with a relatively large field of view of 9.6 square degrees. 

The observing schedule that will be used has not yet been finalized, however, and is in many ways still open to suggestions for how to best optimize LSST. At this point, LSST observations can be quantized as individual 'visits', consisting of two consecutive 34-second exposures in the same band and location, but this could be changed if a compelling science case was presented.
%{\bf is there a citation we can include for this statement?} - This was a statement made at the LSST workshop; I'm not sure how that can be cited, as such
The survey can broadly be divided into a few distinct subsurveys, with the primary 'wide and shallow' survey being what we refer to in this paper as regular cadence fields. These fields will have $\sim$1000 observations over the 10 years of LSST's mission in 6 bands. The other large component we discuss is the 'narrow and deep' survey, or what we refer to as the deep-drilling fields. These will be $\sim$10 fields that will each be observed $\sim$10,000 times in 6 bands, primarily consisting of 40 one-hour block observations for each field. The LSST survey also includes subsurveys that will focus on the galactic plane, polar cap, and the North Ecliptic Spur.

\begin{figure}[!htb]
  \begin{center}
    \begin{subfigure}[b]{0.7\textwidth}
      \includegraphics[width=\textwidth]{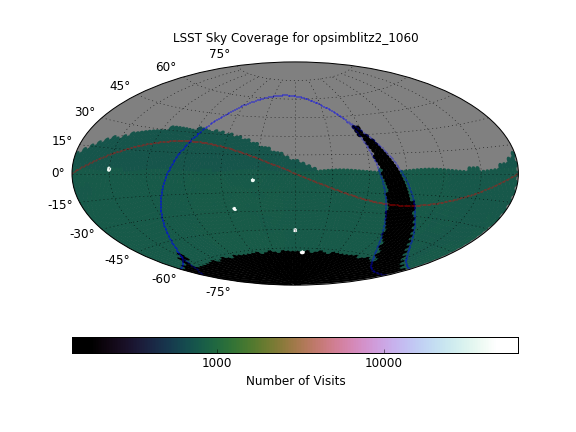}
    \end{subfigure}
    \begin{subfigure}[b]{0.7\textwidth}
      \includegraphics[width=\textwidth]{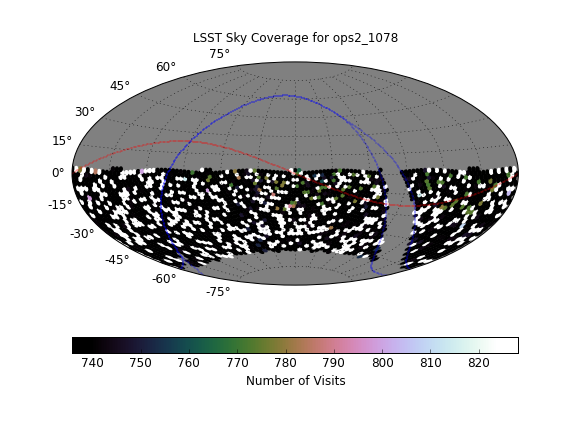}
    \end{subfigure}
  \end{center}
  \caption{An example of a metric as displayed by the Metric Analysis Framework, the total number of observations per field. Opsimblitz2\_1060 on the top, and ops2\_1078 on the bottom. Opsimblitz2\_1060 shows the primary survey, in addition to the deep-drilling fields and additional subsurveys. Ops2\_1078 only features the primary survey.}
  \label{fig:NVisits}
\end{figure}

In this paper we utilize two tools made available by the LSST team: The Operations Simulator (OpSim) and the Metrics Analysis Framework (MAF).
For our purposes, OpSim produces a list of observation times with field ID, band, and limiting magnitude. For this paper we use two OpSim data sets; %\kgsins{[why these specifically?]};
opsimblitz2\_1060 (also referred to as ob2\_1060) consists of several subsurveys and has been introduced as a potential reference run, and ops2\_1078 only consists of the 'wide and shallow' regular cadence fields, but with different cadence parameters\footnote{Available at \url{https://confluence.lsstcorp.org/display/SIM/}}. The number of observations for both fields are shown in Figure~\ref{fig:NVisits}.

Figure~\ref{fig:NVisits} is the output produced by the MAF for a simple metric that calculates the number of observations per field. The LSST team has already provided several sample metrics as part of the initial MAF package.  However, third parties can also create independent metrics. This capability allows for the community to provide new metrics to address properties of the survey tied to the observability of a myriad of scientifically interesting objects.
In what follows, we use the LSST MAF to demonstrate three new metrics that permit quantitative analysis of the performance of different OpSim designs for the survey cadence and coverage, specifically focused on optimizing LSST's coverage of different stellar populations, variables, and transients.

\section{New Metrics for Stellar Populations, Variables, and Transients}\label{sec:metrics}
\subsection{Observation Triplets: Nonrecurring stellar transients}\label{sec:triplets}

A basic concern for LSST centers on its ability to detect transient events. However, many of the metrics that have already been produced for the Metric Analysis Framework focus on coadded images rather than considering how to detect time-dependent photometric variability. This broad category of potential transient events ranges from asteroid collisions to stellar novae and stellar mergers. As all these events require observations during a fixed time interval, an important question for LSST will be how survey cadences will impact LSST's ability to detect events that may take place on timescales from hours to days.

For LSST to suitably detect transient events, we need to define the time resolution necessary to detect a given event. Here, we consider an event that consists of a relatively constant baseline flux, followed by a change to a different flux which remains relatively constant for a period of time. An example of such a transient event is a nova that brightens after a long time of constant brightness, without any simple periodicity. In order to consider this a detection, we require one observation prior to the event, a second observation at the start of the event showing the change in brightness, and a third observation that confirms that the change in brightness wasn't due to noise in a single data point. From these three data points, the magnitude change can be measured and the time of the event can be constrained between the first and second observations, allowing for basic detection and characterization. We can then quantify the ability of a given LSST field to detect non-periodic events by counting the number of these "observation triplets" that occur. 

\begin{figure}[!htb]
  \begin{center}
   \includegraphics[width=\textwidth]{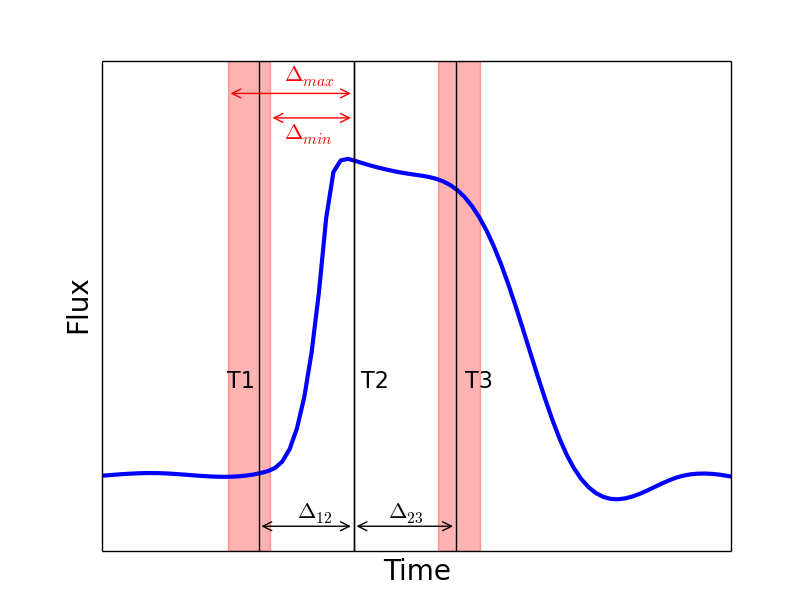}
  \end{center}
  \caption{An illustration of a basic light curve of a non-periodic brightening. T1, T2, T3 represent a set of observations that would satisfy our criteria to detect this event. Using T2 as our reference point, T1 and T3 could fall anywhere within the regions shaded in red that are bounded by a minimum and maxmium time interval.}
  \label{fig:SS_Cyg}
\end{figure}

To provide an example, we look at the case of detecting a nova. In Figure~\ref{fig:SS_Cyg}, we show a simplified light curve of the nova SS Cyg based on AAVSO observations \citep{AAVSO}. The vertical line, T2, marks an ideal time for the middle observation of the observation triplet, shortly after the brightness has increased. We then require that the interval between the first and second, and second and third, observations to fall somewhere between $\Delta_{min}$ and $\Delta_{max}$, and these regions are shaded in red in the figure. Two potential observations are depicted by the vertical lines T1 and T3, with $\Delta_{12}$ and $\Delta_{23}$ being the time differences between the first and second, and second and third observations, respectively. We can also put limits on the ratio of these two time intervals.

This metric is written so that $\Delta_{min}$, $\Delta_{max}$, and the maximum ratio of $\Delta_{12}$ and $\Delta_{23}$ are parameters of the metric. For our example, we set the minimum time interval between observations at 9 days, the maximum at 11 days, and the two time intervals must be within 10\% of one another. In essence, this asks how many opportunities LSST will have to observe non-repeating brightening or dimming events that occur on various timescales. 
Note that the nova example is intended to be illustrative only; the point here is to have a metric that captures the sensitivity to transient behavior on various timescales, which the user can modify as needed for the timescale(s) and behavior(s) of interest.
We also require that the observations of a given triplet all be made in the same band. 

\begin{figure}[!htb]
  \begin{center}
    \begin{subfigure}[b]{0.72\textwidth}
      \includegraphics[width=\textwidth]{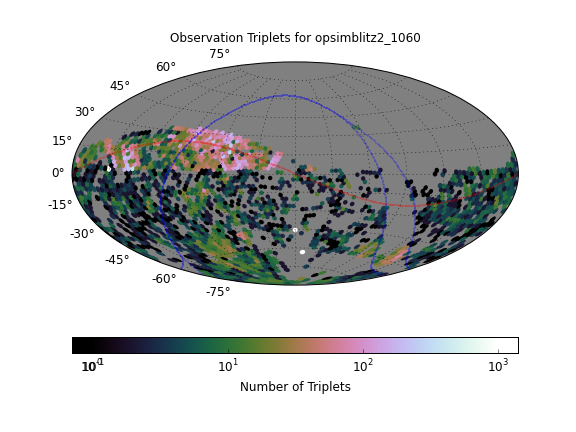}
    \end{subfigure}
    \begin{subfigure}[b]{0.72\textwidth}
      \includegraphics[width=\textwidth]{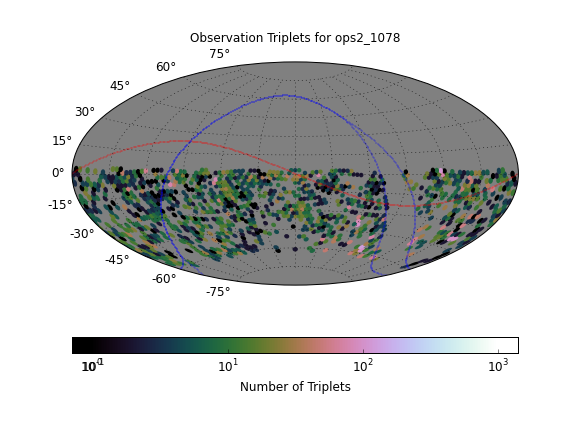}
    \end{subfigure}
  \end{center}
  \caption{The number of observation triplets on an example time scale, with opsimblitz2\_1060 on the top, and ops2\_1078 on the bottom. The observation triplets are more numerous in fields in the North Elliptic Spur and deep-drilling fields shown in opsimblitz2\_1060.}
  \label{fig:Triplet}
\end{figure}

Figure~\ref{fig:Triplet} shows the results we get for two different OPSIM runs. In the first OpSim run the deep drilling fields show up strongly with the number of observation triplets in these fields exceeding 100,000. Additionally, the subsurvey that covers the North Ecliptic Spur, a 4,000 degree area along the ecliptic in the northern hemisphere that has larger airmasses than the main survey observations, also has more observation triplets than the primary wide and fast survey, as do portions of the galactic plane. In the second OpSim, the different cadence results in the number of observation triplets in the main survey being slightly higher, although gaining less than an order of magnitude per field. For the purposes of detecting the novae in this example case, the cadences used in the deep drilling fields, North Ecliptic Spur, and portions of the galactic plane will be far more effective than the cadence of the main survey. The majority of the sky will be surveyed at the regular 'wide and fast' cadence, however, and in the main survey the cadence that is used in OpSim results ops2\_1078 is slightly more beneficial for detecting novae.

\subsection{Periodogram Purity Function: Detectability of Periodic Variability}\label{sec:ppf}

A basic question for LSST is how robustly it can detect periodic behavior. In particular, choosing a cadence for this sort of detection requires some special consideration as the ideal cadence will be one that minimizes the effect of aliasing. To address this, we directly investigate aliasing as caused by the LSST cadence.

LSST will observe numerous kinds of transient periodic objects, such as variable stars, eclipsing binaries, and transiting exoplanets. While each of these kinds of events will have their own unique characteristics, we opt to establish a metric that examines the completeness of phase coverage for a range of periodic timescales, as a way of quantifying LSST's relative likelihood to permit detection of any given periodic variable. 
Specifically, the metric calculates the Fourier power spectral window function of each field \citep{Roberts1987}. The spectral window function provides a measure of how the finite and discrete cadence and sampling of a real light curve result in the Fourier periodogram power getting distributed across multiple alias and beat frequencies. For 
any real light curve, the resulting Fourier periodogram is a combination of the idealized, ``pure" periodogram (i.e., a delta function at one frequency in the case of simple sinusoidal variations) and the spectral window function which ``leaks" that pure power into many other periodogram peaks, decreasing the power of the true frequencies and decreasing the ``purity" of the overall power spectrum. 
%\kgsins{[Keivan to reword preceding sentences a bit.]}

To create our periodogram purity function metric, we calculate the Fourier spectral window function , $F_w$, for a given light curve sampling pattern, and we then 
define the periodogram purity function as $1-F_w$. In the ideal case with no gaps in observation, the spectral window function can be represented as a normalized Kronecker delta centered at 0 and having a height of 1, meaning that all the Fourier power is contained precisely at the correct period and no power is leaked to other frequencies. The periodogram purity function would in this ideal case equal 1 at all nonzero frequencies. As phase coverage is reduced from the ideal case to the real case of finite and discrete data, signal strength will be transferred to other frequencies, creating greater structure in the periodogram. In order to quantify the amount of signal being lost for a given set of observations as a single number for the LSST MAF, we look for the smallest value of the periodogram purity function away from the true frequency. Under that convention, this means that two sets of observations could be compared where the smaller the minimum value of the periodogram purity function is, the more power that has been lost from the actual period. That is, the less ``pure" is the periodogram at frequencies away from the true frequency.

To demonstrate the value of the periodogram purity function when reduced to a single value, we create two sample light curves. In both cases, we begin with a sine curve with a period of 4.5 days. To choose our observing times, we set 1000 evenly spaced observations over the course of ten years. We then add a random number to each observation time in order to have some variation in observing time intervals; in one case we add random values between 0 and 1 days (referred to hereafter as our wide distribution), and in the other case we add random values between 0 and 0.001 days (referred to hereafter as our narrow distribution). In the wide distribution, the time between observations ranges from less than 3 to over 4.5 days. In the narrow distribution, the time between observations ranges only from 3.645 to 3.665.

\begin{figure}[!htb]
  \begin{center}
    \begin{subfigure}[b]{0.45\textwidth}
      \includegraphics[width=\textwidth]{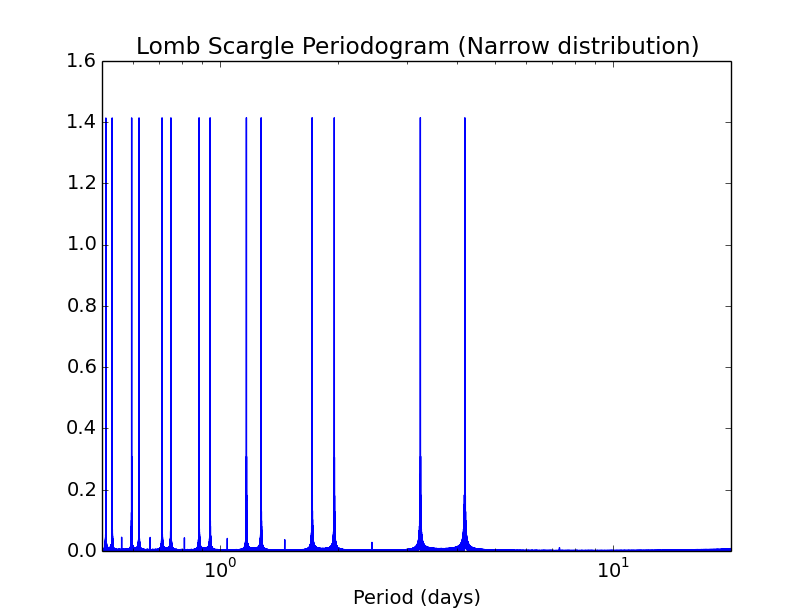}
    \end{subfigure}
    \begin{subfigure}[b]{0.45\textwidth}
      \includegraphics[width=\textwidth]{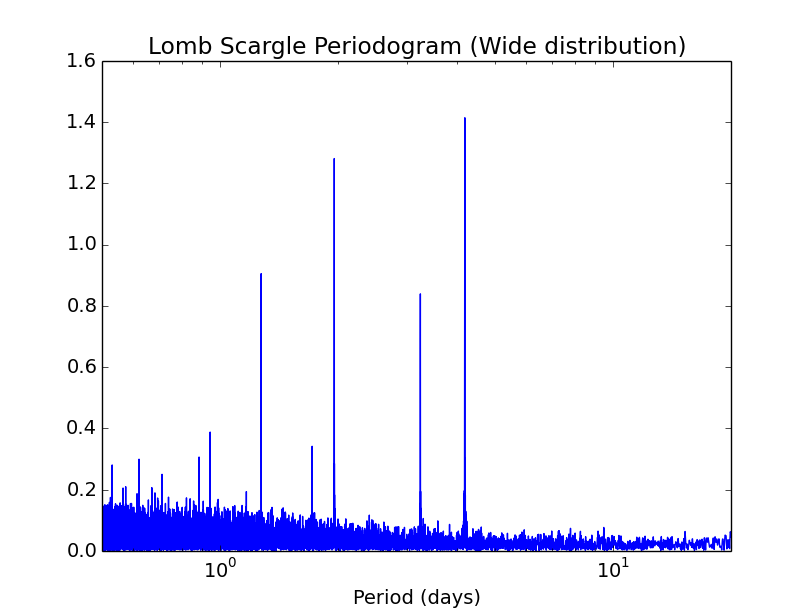}
    \end{subfigure}
    \begin{subfigure}[b]{0.45\textwidth}
      \includegraphics[width=\textwidth]{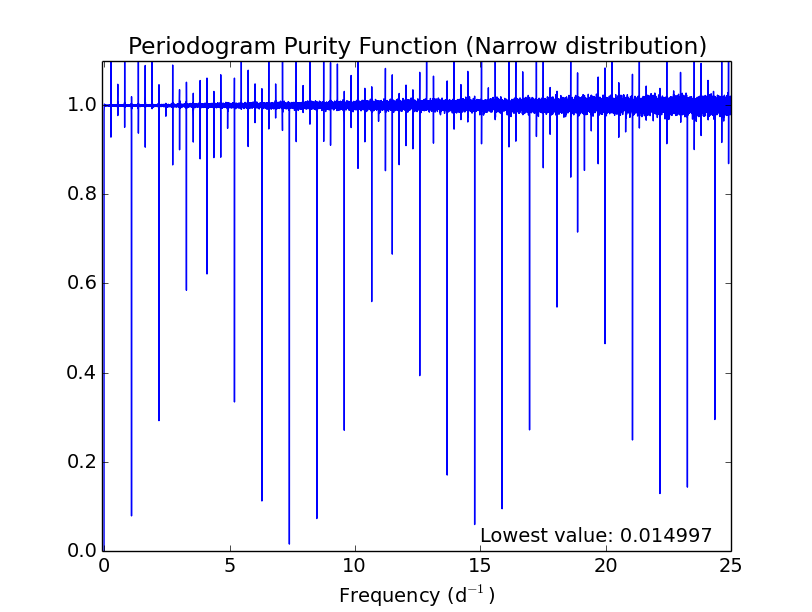}
    \end{subfigure}
    \begin{subfigure}[b]{0.45\textwidth}
      \includegraphics[width=\textwidth]{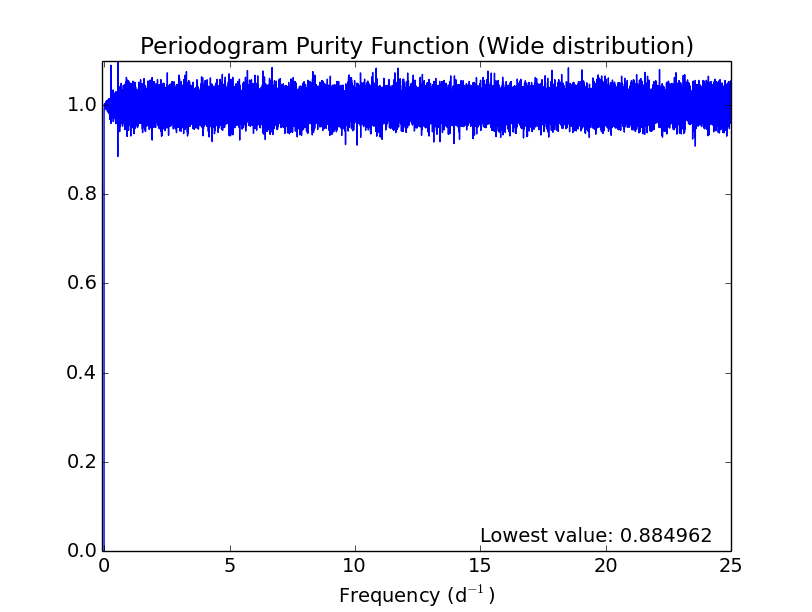}
    \end{subfigure}
  \end{center}
  \caption{The top row shows the Lomb-Scargle periodograms for the two sample light curves described in \S \ref{sec:ppf}. The bottom row shows the corresponding periodogram purity functions. }
  \label{fig:Periodogram}
\end{figure}

To compare these two light curves, we calculate the Lomb-Scargle periodogram\citep{Lomb1976, Scargle1982} and the corresponding periodogram purity function, and these are displayed in Figure~\ref{fig:Periodogram}. The periodogram for the narrow distribution shows numerous peaks of comparable power, with 13 peaks within 1\% of the strength of the top peak. In contrast, for the wide distribution the highest peak correctly corresponds to the period of the sine curve we added to the light curve. Qualitatively speaking in this case, the wide distribution appears to be much better at determining the period of the light curve. When we compare the periodogram purity function, we see that for the narrow distribution that poorly recovered the period, the lowest peak at nonzero frequency shift is less than 0.02. The wide distribution that does recover the period correctly has its lowest peak at just over 0.91, demonstrating that the light curve that was able to more definitively identify the period and have greater phase coverage was also the one with the periodogram purity function value much closer to 1. These periodogram purity function values are agnostic with respect to the nature of the periodic signal we are interested in. An important note is that the spectral window function, and in turn the periodogram purity function, represents how power is distributed to other frequencies from the true period. Changes to the periodic function we are interested in, such as variations in the amplitude or different functional forms, will impact the power that is present; however, the redistribution of this power can be represented by the same term for a fixed set of observation times. 

\begin{figure}[!htb]
  \begin{center}
    \begin{subfigure}[b]{0.45\textwidth}
      \includegraphics[width=\textwidth]{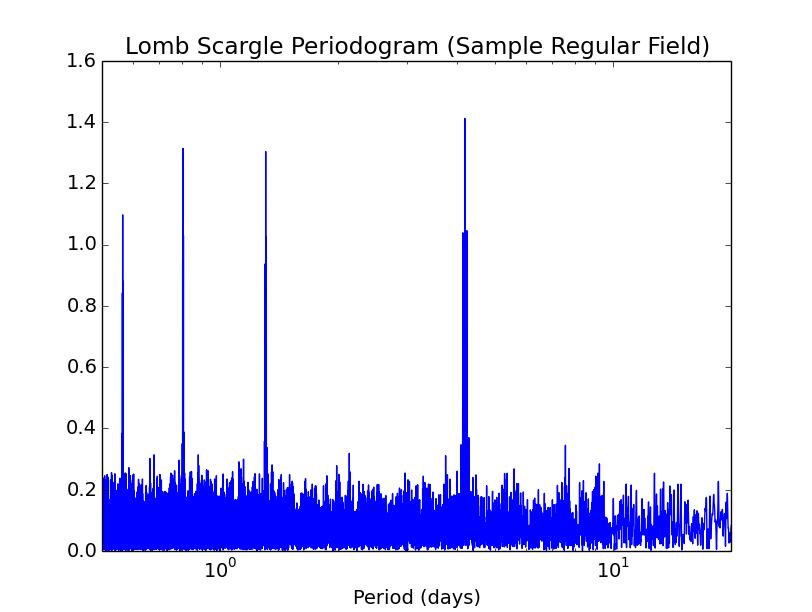}
    \end{subfigure}
    \begin{subfigure}[b]{0.45\textwidth}
      \includegraphics[width=\textwidth]{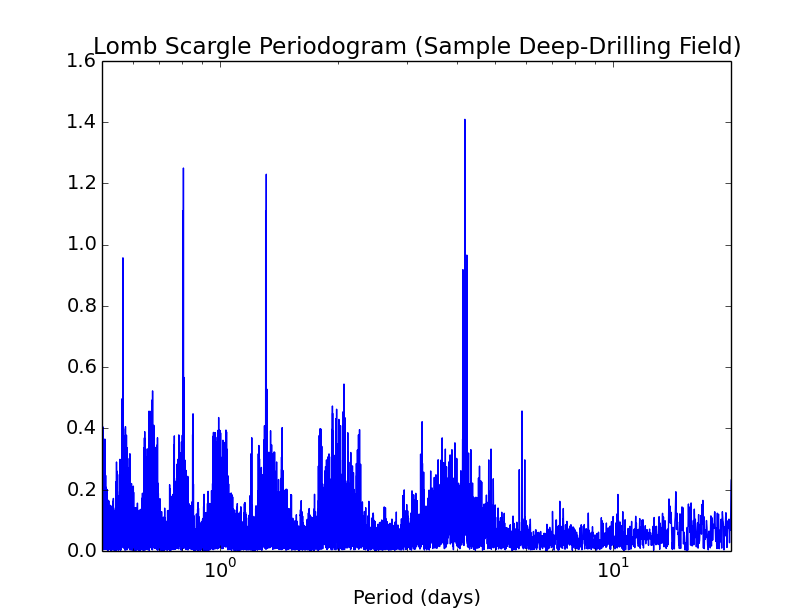}
    \end{subfigure}
    \begin{subfigure}[b]{0.45\textwidth}
      \includegraphics[width=\textwidth]{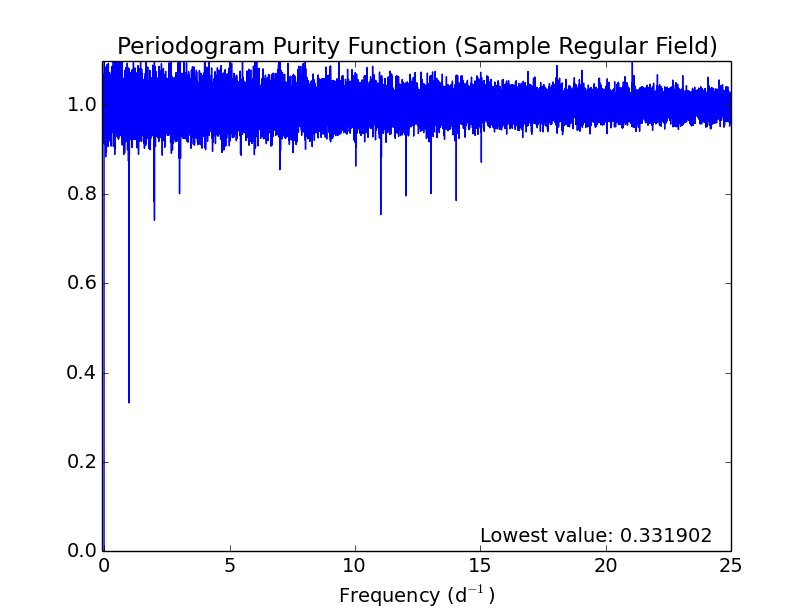}
    \end{subfigure}
    \begin{subfigure}[b]{0.45\textwidth}
      \includegraphics[width=\textwidth]{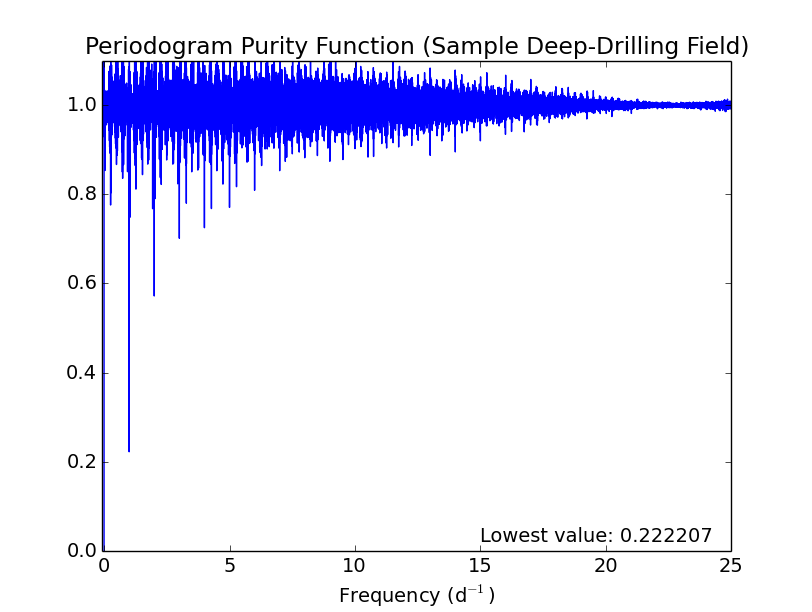}
    \end{subfigure}
  \end{center}
  \caption{The top row shows the Lomb-Scargle periodograms for two fields from the OpSim results; on the left is a regular field and on the right is a deep-drilling field. The bottom row shows the corresponding periodogram purity functions.}
  \label{fig:OpsimPeriodogram}
\end{figure}
We can build upon the sample cadences in Figure~\ref{fig:Periodogram} by looking at two actual cadences from the OpSim results. We choose a regular field consisting of on the left in Figure~\ref{fig:OpsimPeriodogram}, and a deep-drilling field on the right. It is worthwhile to note that for the metric we take the lowest value at a non-zero frequency for the value of the periodogram purity function, and there is very little difference in this value between a regular field and a deep-drilling field. However, unlike our initial test cases, there is interesting non-uniform structure found in the periodogram purity functions, with the deep-drilling fields having more scatter at lower frequency shifts, indicating that there is more loss at small changes in frequency, and we can expect to see broader peaks in the periodogram due to this.
\begin{figure}[!htb]
  \begin{center}
    \begin{subfigure}[b]{0.7\textwidth}
      \includegraphics[width=\textwidth]{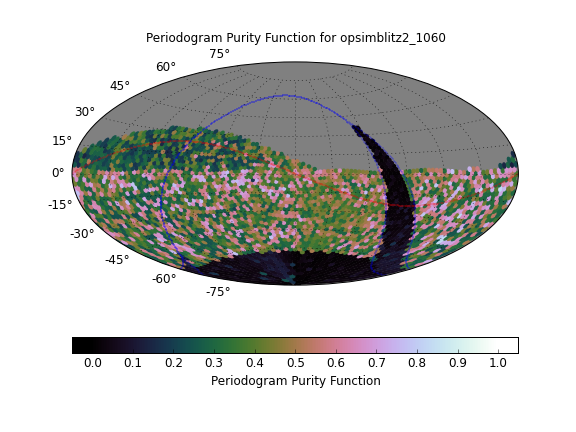}
    \end{subfigure}
    \begin{subfigure}[b]{0.7\textwidth}
      \includegraphics[width=\textwidth]{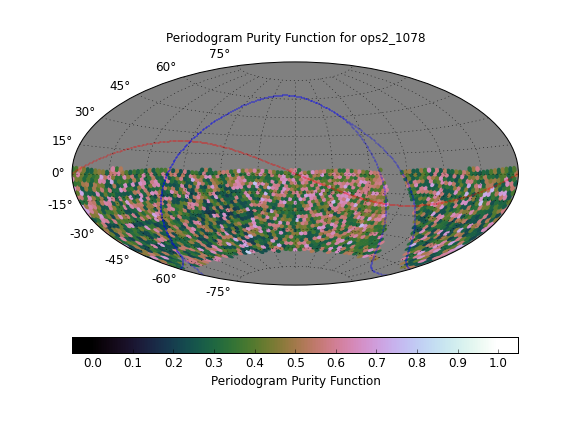}
    \end{subfigure}
  \end{center}
  \caption{Periodogram purity function values for each field in two simulated observing runs, opsimblitz2\_1060 on the top, and ops2\_1078 on the bottom. An ideal time sampling would have a periodogram purity function value of 1; this function is only dependent on the times of observation.}
  \label{fig:SWF_MAF}
\end{figure}

Using the Metric Analysis Framework, we then look at the value from the periodogram purity function for every field in two observing simulations in Figure~\ref{fig:SWF_MAF}. A general characteristic that appears is that in the overall survey, each subsurvey represents a different regime of phase coverage. These results are dependent only on the observing schedule and are not dependent on any input signal. For the first OpSim result, we see that the main survey has the best phase coverage, and the other subsurveys will have worse phase coverage. In the second OpSim result, the adapted cadence seems to have approximately the same phase coverage.

For any observations that will require high quality phase coverage, we see that the south celestial pole and galactic plane do not provide any significant phase coverage. Additionally, the North Ecliptic Spur and the deep-drilling fields will have phase coverage equal to the main survey fields. The current metric we provide for use with the Metric Analysis Framework provides some insight, however there may be further benefit in an increased study of the structures shown in the periodogram in Figure~\ref{fig:OpsimPeriodogram} to develop a more sophisticated understanding of what phases LSST will be able to best recover, particularly in the deep-drilling fields.

\subsection{Field Star Counts}

%\kgsins{A key question for LSST is... [flesh out this intro paragraph a bit... what is the question being asked?]. For example, how many stars per field are suitable for exoplanet transit detection? Cite Lund and Jacklin papers.}
There are two principal components that will need to be considered to understand the total yield of any transient object. The first component is the cadence, which is measured in the metrics presented in Sections \ref{sec:triplets} and \ref{sec:ppf}. The other important factor is the number of potential sources in a given field, which for stellar sources would represent the number of stars that would be observed with sufficient signal to noise to be useful, while also being faint enough that the LSST detector will not saturate. In order to address this concern, we develop a metric to determine the number of stars in a given field that satisfy some set of observation parameters.

\begin{figure}[!htb]
  \begin{center}
   \includegraphics[width=0.9\textwidth]{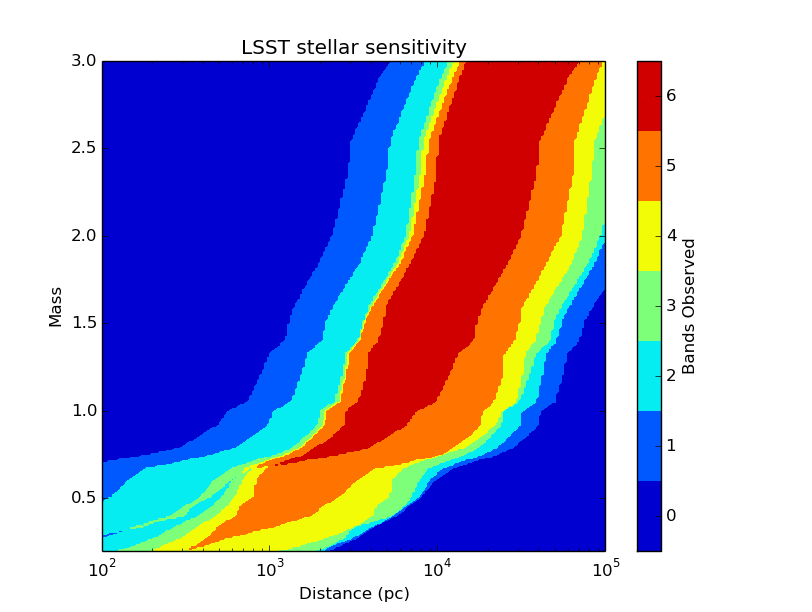}
  \end{center}
  \caption{Number of bands in which a main sequence star will be observable between a saturation-limited upper brightness and a noise-limited lower brightness.}
  \label{fig:Regime}
\end{figure}

LSST will be able to observe a very wide range of stars with relatively low noise, including stars as small as red dwarfs, and distances ranging from within $\sim$100 parsecs out to over 10,000 parsecs, including the LMC. This will be, in part, due to the multiple bands that LSST will observe in. We show this wide range in Figure~\ref{fig:Regime}, where we examine how many bands main sequence stars would be observable in. For our bright limit we use the apparent magnitude of 16 as a saturation point. For our faint limit, we required the photometric noise to be less than 0.03 mag, a value that we used as a cutoff for the maximum noise while still being able to detect a transiting planet in \citet{Lund2014}, or an approximately 1\% drop in brightness during transit. We convert from stellar masses to spectral types by interpolating the relations from \citet{Cox2000}. Based on spectral type, we obtain absolute magnitudes in the \emph{ugriz} bands from \citet{Covey2007}. While there is still much uncertainty about the specific parameters of the \emph{y}-band filter that LSST will eventually use, for this work we use a \emph{y}-band defined by \citet{Hodgkin2009}. This allows us to convert a set of stellar masses into a set of absolute magnitudes.

There are a few notable features of Figure~\ref{fig:Regime}. There are two excluded regions of magnitude/distance where LSST would not be sensitive at the specified thresholds. The empty upper left corner shows that all stars more massive than $\sim1.5 M_{\odot}$ and closer than 1 kpc are saturated in all LSST bands. The lower right corner shows that all stars less massive than $\sim0.7 M_{\odot}$ and more distant than 10 kpc will not be detected with the specified photometric precision for exoplanet transit detection in any LSST band. Running diagonally through the Figure, there is a very large range in stellar masses and distances where stars will be measured with sufficient sensitivity to detect small variation such as transiting planets. Stars with masses below 0.5 $M_\odot$ are observable within $\sim$4000 parsecs in several bands, and as close as $\sim$100 parsecs. Solar-mass stars are observable in at least 5 bands from $\sim$2000 to $\sim$20,000 parsecs, a distance range that includes the galactic bulge. Stars slightly larger than one solar mass that are in the Magellanic Clouds, at $\sim$50,000 parsecs, would be observed within the specified constraints.

As the range of observable stellar masses is extremely broad, there is a great benefit in determining the number of stars that satisfy these parameters for a given type of event. LSST covers a large region of sky, and so the method we use is designed to not be computationally intensive, or to rely on a more thorough and complex method such as the simulation process that is conducted by TRILEGAL \citep{Girardi2005}. Instead, we only calculate the number of stars within a volume based on stellar density, rather than attempt to create a simulated population of stars.

\begin{figure}[!htb]
  \begin{center}
   \includegraphics[width=0.9\textwidth]{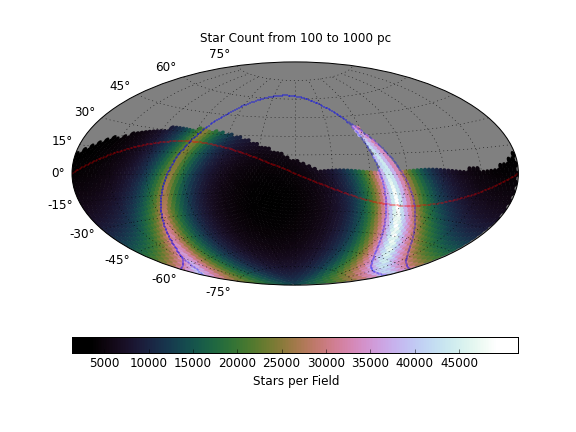}
  \end{center}
  \caption{Stellar count per field between 100 and 1000 parsecs}
  \label{fig:Starcount}
\end{figure}

The basic version of this metric calculates the number of stars within a given field and within a particular distance range. We accomplish this by treating each field as a truncated cone spanning two distances. We then break this cone into sections of equal volume and assign a stellar density for each volume as a function of the galactic coordinates of the center of that section. This begins with the analytical equation for the density of stars in the Galaxy, as a combination of the thin disk, thick disk, and halo, as presented in \citet{Juric2008}. As \citet{Juric2008} had focused only on the local portion of the Milky Way, that equation does not include a contribution from the galactic bulge, and so we include a term for the galactic bulge from \citet{Jackson2002}. Without applying any brightness cut, we determine a stellar count in each field (integrated from 100 to 1000 pc as a demonstration) as shown in Figure~\ref{fig:Starcount}.

\begin{figure}[!htb]
  \begin{center}
   \includegraphics[width=\textwidth]{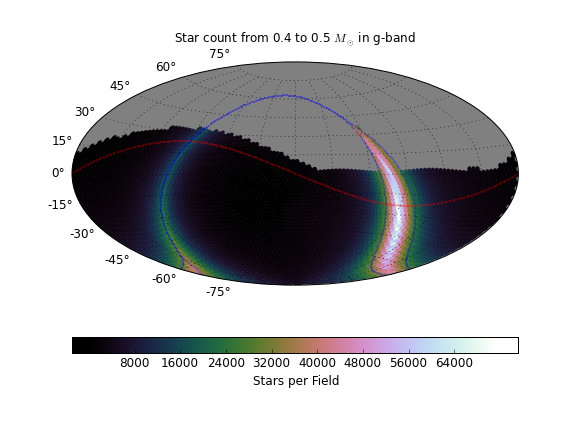}
  \end{center}
  \caption{Star count per field for the 0.4-0.5 solar mass range and noise less than 0.03 mag. Observations in the \emph{g}-band.}
  \label{fig:StarcountMass1}
\end{figure}

\begin{figure}[!htb]
  \begin{center}
   \includegraphics[width=\textwidth]{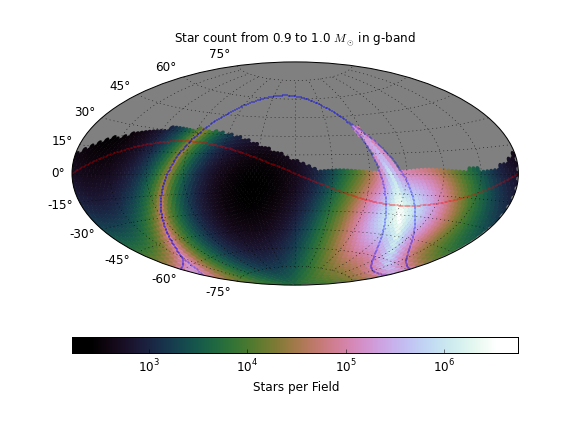}
  \end{center}
  \caption{Star count per field for the 0.9-1.0 solar mass range and noise less than 0.03 mag. Observations in the \emph{u}-band.}
  \label{fig:StarcountMass2}
\end{figure}

To look at specific stellar masses, we define a mass range of interest. This set of stellar masses is then converted into a set of absolute magnitudes. Combining those values, the noise model, and the apparent magnitude limit, we can determine a distance range over which these stars would be within LSST's sensitivity for a set of stellar mass bins. We then calculate the total star count in the distance range that corresponds to each stellar mass bin, and then select only the stars with masses in that stellar mass bin by applying a standard Salpeter IMF. We demonstrate this technique with two example mass ranges, with stars between 0.4 and 0.5 solar masses in the \emph{g} band in Figure~\ref{fig:StarcountMass1}, and stars between 0.9 and 1.0 solar masses in the \emph{u} band in Figure~\ref{fig:StarcountMass2}. The Field Star Count metric can be used in conjunction with metrics that are developed for stellar observations, as a field that performs well in a given metric may not be useful if there are many stars in that particular field.


\section{Summary}\label{sec:sum}

In this paper we have introduced three new metrics to help compare LSST observing strategies in advance of the start of LSST's mission in 2020. The Observation Triplets provide a tool to study any transient event that could be sufficiently discovered or characterized by an observation at a baseline brightness and two observations detecting and confirming the event. The tool can be adapted for a range of timescales, however we provide the specific example of a ten-day timescale that would be suitable for stellar novae. For this example, the deep-drilling fields and the North Ecliptic Spur would be the cadences that are most useful for finding stellar novae.

The Periodogram Purity Function metric is a tool to probe the phase coverage of the LSST fields. The South Celestrial Pole and the galactic plane subsurveys both have very poor phase coverage and would have very limited usage for studying periodic variables. The Periodogram Purity Function metric currently only uses a single value from the Periodogram Purity Function, and there may be some benefit in an effort to try to better characterize the periodogram purity function by using different criteria to pick this single value. Additionally, while a larger periodogram purity function value should result in the period being easier to detect and characterize correctly, it is not readily apparent to what extent an increase in this value will result in an increased likelihood of detection.

In the Field Star Count metric, we provide a tool to quantify the number of stellar sources that will be present in LSST fields. We look at a few limited mass ranges in particular bands, but it provides indications that we can reasonably expect the number of stars around which planets could be detected (not saturated but sufficiently low noise) is a number in the tens of thousands per field for red dwarfs, and for solar-mass stars exceeds 100,000 stars per field in the direction of the galactic bulge. To use the example of transiting planets where hot Jupiters occur around 1\% of stars, and around 10\% of hot Jupiters transit, this represents around 10 transiting hot Jupiters around red dwarfs per field (30,000 total), and exceeding 100 transiting hot Jupiters on 0.9-1.0 M$_{\odot}$ stars per field in the direction of the galactic bulge. This doesn't include detection efficiencies or other regimes of planets, but provides a good example of the information that can be gained from considering the number of stellar sources in LSST.

%\section{Acknowledgements}
\acknowledgments
We thank the attendees of the LSST Observing Cadences Workshop, and the fast transient group led by Mansi Kasliwal and Lucianne Walkowicz in the early development of the Observation Triplets metric. We also are grateful for the assistance from the LSST team in troubleshooting issues with the MAF.

\bibliographystyle{apalike}
%\bibliography{LSST_Ib}
%\bibliographystyle{apalike}
\bibliography{libAAS}

\end{document}